\documentclass[10pt,twocolumn]{article}

\usepackage[top=2.0cm,bottom=2.5cm,left=1.8cm,right=1.8cm,
            columnsep=0.6cm]{geometry}
\usepackage{amsmath,amssymb}
\usepackage{graphicx}
\usepackage{xcolor}
\usepackage{booktabs}
\usepackage{microtype}
\usepackage[hidelinks]{hyperref}
\usepackage{abstract}
\usepackage{times}
\usepackage[T1]{fontenc}
\usepackage{pgfplots}
\usepackage{tikz}

\usetikzlibrary{arrows.meta,positioning,calc,backgrounds,patterns}
\pgfplotsset{compat=1.18}

\usepackage{titlesec}
\titleformat{\section}{\normalfont\bfseries\normalsize}{\thesection.}{0.5em}{}
\titleformat{\subsection}{\normalfont\itshape\normalsize}{\thesubsection.}{0.5em}{}
\titlespacing{\section}{0pt}{8pt plus 2pt minus 1pt}{3pt}
\titlespacing{\subsection}{0pt}{5pt}{2pt}
\usepackage[font=small,labelfont=bf,labelsep=period]{caption}

\definecolor{w7xblue}{RGB}{31,119,180}
\definecolor{lhdorange}{RGB}{214,99,14}
\definecolor{hgreen}{RGB}{44,160,44}
\definecolor{tjred}{RGB}{196,48,43}
\definecolor{mobedge}{RGB}{130,80,180}
\definecolor{extblue}{RGB}{200,225,245}
\definecolor{locgray}{RGB}{215,215,215}
\definecolor{puckgold}{RGB}{200,160,0}

\newcommand{\etai}{\eta_i}
\newcommand{\etastar}{\eta_i^{*}}
\newcommand{\etailin}{\eta_i^{\rm lin}}

\newcommand{\epn}{\epsilon_n}
\newcommand{\Ln}{L_n}
\newcommand{\LT}{L_{T_i}}
\newcommand{\xiL}{\xi}
\newcommand{\lam}{\lambda}

\newcommand{\lamAAH}{\lambda_{\rm AAH}}

\usepackage{afterpage}
\usepackage{placeins}
\usepackage{fancyhdr}
\begin{document}

\twocolumn[{%
\begin{@twocolumnfalse}
\vspace*{4pt}
{\noindent\LARGE\bfseries
  Anderson localization of Ion-Temperature-Gradient Modes 
  in Aperiodic Stellarators\par}
\vspace{10pt}
{\noindent\large Amitava Bhattacharjee$^{1}$\par}
\vspace{4pt}
{\noindent\small
  $^{1}$Department of Astrophysical Sciences, Princeton University\quad  
  Correspondence: \texttt{amitava@princeton.edu}\par}
\vspace{10pt}\hrule height 0.6pt\vspace{8pt}
\begin{abstract}
\noindent
The geometric coefficients of the ion-temperature-gradient (ITG) eigenmode equation along a magnetic field line
are aperiodic because the line samples the full three-dimensional
geometry.
Starting from a reduced fluid model for drift waves
[\textit{Phys.\ Fluids} \textbf{26}, 880 (1983)], we show that
aperiodic stellarator geometry leads to a quasiperiodic Hill equation
for the ITG mode structure. In a tight-binding
approximation this equation reduces to an Aubry--Andr\'{e}--Harper
difference equation, suggesting an Anderson-localization mechanism
for  ITG eigenfunctions.
For the continuous equation the localization threshold $\etastar$ is
identified by the onset of a positive Lyapunov exponent of the transfer
matrix at the drift-wave-resonant eigenvalue. The result is a structural property of
the eigenvalue problem and is obtained on the drift-wave-resonant branch
of the fluid model, at weak global and local magnetic shear. We discuss
what bearing, if any, this transition may have on the observed clamping of
the ion temperature in stellarators, a question that is not settled by this theory.
\end{abstract}
\vspace{8pt}\hrule height 0.6pt\vspace{14pt}
\end{@twocolumnfalse}
}]

\section{Introduction}
\label{sec:intro}

A persistent feature of electron-heated stellarator experiments is the saturation of ion temperature despite increasing heating power. In Wendelstein 7-X (W7-X)~\cite{Beidler2021,Beurskens2021} and the Large Helical Device (LHD)~\cite{Nagaoka2015} the ion temperature $T_i$ typically remains fixed, even when the applied heating power is increased substantially and the electron temperature $T_e$ increases over a wide range. This phenomenon is commonly referred to as \emph{ion-temperature clamping}. 
Beurskens et al.~\cite{Beurskens2021} attribute the effect to a combination of
the limited and broad ion heating supplied by electron--ion energy exchange in
electron-cyclotron-heated plasmas, the gyro-Bohm increase of the heat
diffusivity, and the destabilising influence of a rising $T_e/T_i$ on
ion-temperature-gradient (ITG) turbulence. Standard quasilinear ITG
transport theory, by contrast, would have the ion temperature rise with
deposited power~\cite{Romanelli1989,Romanelli1993}. Two features of the observation
should be noted at the outset. The clamping is reported
to be largely independent of magnetic configuration~\cite{Beurskens2021},
and a closely similar saturation is seen in $L$-mode plasmas of the
axisymmetric device ASDEX Upgrade~\cite{Beurskens2022}. A mechanism
resting on three-dimensional geometry (3D) alone is unlikely, therefore, to be the whole
explanation of clamping, and we do not claim so here.
Furthermore, power-balance studies at W7-X show substantial scatter in
$a/\LT$ at a given $a/\Ln$~\cite{Wappl2025}, so we do not make the claim that the drive parameter
$\etai = \Ln/\LT$, the dimensionless ratio of the logarithmic density
gradient to the logarithmic ITG scale lengths, locks at a
fixed value. In what follows, we focus instead on a structural property of the
ITG eigenvalue problem in 3D geometry which can influence a
phenomenological account of profile data.

Specifically, we investigate the Anderson localization~\cite{Anderson1958} of ITG eigenmodes caused by the 3D magnetic geometry. In the ballooning representation of the ITG instability, one obtains a differential equation along the magnetic field line whose coefficients depend on the magnetic curvature and drift frequencies. In stellarators these quantities are not strictly periodic along a field line but instead vary aperiodically because the field line samples the full 3D geometry of the magnetic field. As a result, under some simplifying assumptions, the ITG eigenvalue problem becomes an \emph{aperiodic Hill equation}. Differential operators of this type are closely related to the Aubry--Andr\'{e}--Harper (AAH) model~\cite{Aubry1980,Harper1955}. Such operators are known to exhibit localization transitions in which eigenfunctions change from spatially extended Bloch states to exponentially localized Anderson states as the amplitude of the aperiodic modulation increases. Such a phenomenon has been recently identified in the context of linear~\cite{CuthbertDewar2000,RediJohnson2002} and nonlinear magnetohydrodynamic (MHD) stability~\cite{Bhatt2026} of ballooning modes in stellarators.

The Hill-equation structure of the ITG problem in W7-X geometry was
recognized by Zocco et al.~\cite{Zocco2018}, who obtained the
destabilization threshold and found
extended Floquet-type modes to persist in W7-X. Such modes were
subsequently identified in gyrokinetic simulations near
marginality~\cite{Podavini2024}. That work solves the ITG eigenvalue problem
itself for a periodic Hill equation. The present paper asks a different
question of the same operator --- what becomes of the mode structure when
the coefficients are aperiodic rather than periodic --- and does so on
the reduced drift-wave-resonant branch.

The key idea of the present paper is that an Anderson localization transition can occur for ITG eigenmodes in stellarators. The discrete AAH limit has a sharp localization transition with no mobility edge, that is, no energy within the spectrum separating extended from localized states: all states localize together at a single value of the coupling (in condensed matter physics parlance). The continuous quasiperiodic Hill equation considered here is AAH-like, but its threshold must be computed from the Lyapunov exponent at the relevant eigenvalue. We therefore use the onset of a positive Lyapunov exponent at the drift-wave-resonant energy as the localization diagnostic for the reduced fluid model.

The localization threshold $\etastar$ depends on the dimensionless magnetic drift and curvature profile along the field line. The precise threshold for any real 3D device requires the full Boozer harmonic spectrum for the background equilibrium, which is sensitive to geometry. The transition point $\etastar$ marks the boundary between an extended regime ($\etailin < \etai < \etastar$) and an Anderson-localized regime ($\etai > \etastar$). Here $\etailin$
denotes the linear ITG instability threshold, the value of $\etai$ above
which the mode is linearly unstable. It is not determined by the model used
here, which is a mode-structure equation rather than a dispersion relation
(Section~\ref{ssec:bhatt}). What this change of mode
structure implies for cross-field transport is a separate question, on which
arguments run in both directions; we set them out in
Section~\ref{ssec:transport}.

If the proposed localization mechanism is confirmed by simulation and experiment, it could establish a direct connection between stellarator magnetic geometry and Anderson localization physics, providing a new perspective on turbulent transport in magnetically confined plasmas.

\section{From the two-fluid ITG equation to the Aubry--Andr\'{e}--Harper map}
\label{sec:model}

\subsection{The two-fluid ITG equation and its Schr\"{o}dinger form}
\label{ssec:bhatt}

In the long-wavelength fluid limit, with the ballooning
representation~\cite{Connor1978} and a \emph{straight stellarator} geometry,
Bhattacharjee et al.~\cite{Bhattacharjee1983} reduced the drift-wave
eigenmode problem to a single Schr\"odinger-like equation for the
normalized electrostatic potential $\phi \equiv e\tilde{\phi}/T_e$
along the field-line coordinate $\theta = k_0 s$ (where $s$ is arc
length along the field line and $k_0 = \omega_{*e}/c_s$ is the
normalization wavenumber with $\omega_{*e}$ is the electron diamagnetic frequency, $c_s^2 \equiv T_e/m_i$, and $m_i$ is the ion mass).
Here we extend their treatment by including an ion-temperature gradient. The relevant two-fluid ITG equation is derived in
Appendix A:
\begin{equation}
    \frac{d^2\phi}{d\theta^2}
    +
    \left[
        \Omega(\Omega-1)
        +
        \epsilon_d(\theta)(\Omega+\eta_i)
    \right]\phi
    =
    0.
    \label{eq:parent}
\end{equation}
Here $\Omega \equiv {\omega}/{\omega_{*e}}$ is the normalized eigenfrequency, \(\epsilon_d(\theta)=\epsilon_n K(\theta)\), $\epn = \Ln/R$ is the ratio of density gradient scale length to major radius, $\eta_i = L_n/L_{T_i}$ is the dimensionless ratio of the logarithmic density gradient scale length to the logarithmic ion-temperature-gradient scale length. We define the quantity $\omega_d(\theta)$ as the combined magnetic curvature and gradient-$B$ drift frequency, defined through the field-line variation of the magnetic field strength:
\begin{equation}
  \omega_d(\theta)
  \;\equiv\; \omega_{*e}\,\epn\,K(\theta)= \omega_{*e}\epsilon_d(\theta)\,
  \label{eq:omegad}
\end{equation}
Here the normalization by $\epn$ ensures $|K|\sim 1$. We choose model forms for  $K(\theta)$ to represent tokamaks and stellarators qualitatively in this paper.

We now consider the drift-wave-resonant branch
\[
\Omega=1+\Lambda,\qquad |\Lambda|\ll 1 .
\]
Then
\[
\Omega(\Omega-1)=\Lambda+O(\Lambda^2),
\qquad
\Omega+\eta_i=1+\eta_i+O(\Lambda).
\]
We then obtain the Hill equation
\begin{equation}
\frac{d^2\phi}{d\theta^2}
+
\epsilon_n(1+\eta_i)K(\theta)\phi
=
-\Lambda\phi .
\label{eq:hill}
\end{equation}

We emphasize that this reduced equation is a leading-order
mode-structure equation. For real \(K(\theta)\), it
does not by itself determine the ITG growth rate. Growth rates and a
quantitative critical gradient require the effects omitted here, including
finite-Larmor-radius corrections, kinetic resonances, non-adiabatic ion
response, and the full stellarator magnetic-drift geometry. The expansion about
$\Omega = 1$ selects the drift-wave-resonant branch of
Eq.~\eqref{eq:parent}. It does \emph{not} describe the ITG branch proper,
whose frequency is not close to $\omega_{*e}$ (the linearization discards
the second root of the quadratic in $\Omega$). What is computed in
Section~\ref{sec:clamp} is therefore the localization of the mode
structure of the reduced operator evaluated at $\Lambda = 0$, and not the
localization of the ITG eigenmode at its own eigenfrequency. We expect the
transition to survive that extension, because the Lyapunov exponent of the
transfer matrix is defined for complex $\Lambda$ just as it is for real
$\Lambda$, and the aperiodicity of the coefficients --- which is what
drives the localization --- is a property of the geometry rather than of
the branch.

The normalization $\Omega = \omega/\omega_{*e}$ and the entire
framework that follows are valid only when the drift-wave frequency
$\omega_{*e}$ is finite and non-zero, i.e.\
when the density gradient is finite ($L_n < \infty$, $\etai$
well-defined). In the flat-density limit $L_n \to \infty$ (equivalently $a/L_n \to 0$,
$\etai \to \infty$), the drift-wave frequency vanishes, and
the ITG mode becomes a purely curvature-driven mode not described by
Eq.~\eqref{eq:parent}. Gyrokinetic simulations performed at $a/L_n = 0$ (flat density,
$\etai = \infty$) operate in this different regime and are not directly comparable to the results of this paper.
For recent comprehensive numerical studies relevant to W7-X, see~\cite{Podavini2024,Thienpondt2025}.

We note that the reduced equation requires $|\Lambda| \ll 1$ and $\epn \ll 1$,
which the expansion leading to Eq.~\eqref{eq:hill} uses in the combination
$A = \epn(1+\etai)$. The relevant restriction is therefore not on $\etai$
alone but on $A$: the drift-wave-resonant description holds while
$A \lesssim 1$. With $\epn = 0.12$ this permits
$\etai$ up to roughly $7$, which comfortably contains the thresholds
computed below ($\etastar \approx 2$--$3$). The genuine breakdown is the
limit $\Ln \to \infty$ at fixed $\LT$, where $\omega_{*e} \to 0$ and the
normalization itself fails; that limit is approached with $\epn \to 0$ and
$\etai \to \infty$ such that $A$ remains finite, and it is not the limit in
which $\etastar$ is evaluated here.

\begin{figure*}[!t]
\centering
\begin{tikzpicture}[font=\small, >=Stealth]

\begin{scope}[xshift=0cm]
\begin{axis}[
  width=5.8cm, height=5.0cm,
  xlabel={Ballooning angle $\theta/\pi$},
  ylabel={Potential $V(\theta)/\epn(1{+}\etai)$},
  xmin=-3, xmax=3, ymin=-2.2, ymax=2.2,
  xtick={-3,-2,-1,0,1,2,3},
  ytick={-2,-1,0,1,2},
  axis lines=left,
  tick label style={font=\scriptsize},
  label style={font=\scriptsize},
  legend pos=south east,
  legend style={font=\scriptsize,draw=none,fill=white,
                inner sep=2pt,row sep=-2pt},
  title={\textbf{(a)} Effective potential},
  title style={font=\small\bfseries,at={(0.5,1.03)}},
]
  \addplot[thick,w7xblue,domain=-3:3,samples=200]
    {-cos(deg(pi*x))};
  \addplot[thick,tjred,domain=-3:3,samples=400]
    {-(cos(deg(pi*x)) + 0.8*cos(deg(1.618*pi*x)))};

\end{axis}

\end{scope}

\begin{scope}[xshift=6.8cm]
\begin{axis}[
  width=5.8cm, height=5.0cm,
  xlabel={Ballooning angle $\theta/\pi$},
  ylabel={$|\phi(\theta)|$ [arb.\ units]},
  xmin=-4, xmax=4, ymin=0, ymax=1.45,
  xtick={-4,-2,0,2,4},
  ytick={0,0.5,1.0},
  axis lines=left,
  tick label style={font=\scriptsize},
  label style={font=\scriptsize},
  legend pos=north east,
  legend style={font=\scriptsize,draw=none,fill=white,
                inner sep=2pt,row sep=-2pt},
  title={\textbf{(b)} ITG eigenfunction},
  title style={font=\small\bfseries,at={(0.5,1.03)}},
]
  \addplot[thick,w7xblue,domain=-4:4,samples=300]
    {0.85*exp(-0.04*x^2)*abs(sin(deg(pi*x+0.3)))+0.05};
  \addlegendentry{\scriptsize $\lamAAH{<}1$: extended};
  \addplot[thick,tjred,domain=-4:4,samples=300]
    {1.10*exp(-0.55*x^2)*abs(sin(deg(pi*x+0.3)))};
  \addlegendentry{\scriptsize $\lamAAH{>}1$: localized};
  \draw[<->,thick,mobedge]
    (axis cs:-1.35,1.28)--(axis cs:1.35,1.28)
    node[midway,above,font=\tiny,mobedge]{$2\xiL$};
\end{axis}
\end{scope}

\end{tikzpicture}

\caption{
  (a)~Effective potential $V(\theta)$ seen by a mode travelling along a
  single field line: the periodic potential $(-\cos\theta)$ (blue) and the aperiodic potential ($-\cos\theta{-}\lam\cos(\alpha\theta)$) with $\lam=0.8$,~$\alpha=1.618$ (golden ratio) in red. The two-harmonic form is illustrative; a real curvature spectrum contains many harmonics. (b)~Schematic of the ITG eigenfunction $|\phi(\theta)|$ in ballooning space: extended for $\lamAAH<1$ (blue), exponentially localized for $\lamAAH>1$ (red) with localization length $\xiL = 1/\ln\lamAAH$. The criterion involves the effective tight-binding coupling $\lamAAH$ of Appendix~B, not the geometric amplitude ratio $\lam$; the two coincide only in the bare tight-binding limit.}

\label{fig:bhatt_aah}

\end{figure*}

\subsection{Aperiodic stellarator geometry: a simple model}
\label{ssec:aperiodic}
In an aperiodic stellarator the magnetic field strength along a field
line cannot be written as a function of a single angle, because the
field line samples the full 3D geometry.
It may be written instead as a sum of incommensurate harmonics,
\begin{equation}
  B(\theta) = B_0\bigl[1 - B_1\cos\theta - B_2\cos(\alpha\theta+\delta)
              + \cdots\bigr],
  \label{eq:Btheta}
\end{equation}
where $B_1$ and $B_2$ are the amplitudes of the two dominant
oscillatory harmonics along the line, $\alpha$ the ratio of their
field-line wavenumbers, and $\delta$ their relative phase.

For the illustrative model used here, we approximate the magnetic drift function
rather than the magnetic-field strength itself by a two-harmonic form,
\[
    K(\theta)=\cos\theta+\lambda\cos(\alpha\theta+\delta).
\]
In a quantitative stellarator calculation, \(\lambda\) and \(\alpha\) should
be extracted from the Fourier spectrum of \(\omega_d(\theta)\).

For a two-harmonic stellarator model, the reduced equation then becomes
\begin{equation}
\frac{d^2\phi}{d\theta^2}
+
\epsilon_n(1+\eta_i)
\left[
\cos\theta+\lambda\cos(\alpha\theta+\delta)
\right]\phi
=
-\Lambda\phi .
\label{eq:bhatt}
\end{equation}

In the equivalent Schr\"odinger-like equation, the effective potential $V(\theta)$ in Eq.~\eqref{eq:bhatt} is therefore
\begin{equation}
  V(\theta) = -\epn(1+\etai)
    \bigl[\cos\theta + \lam\cos(\alpha\theta+\delta)\bigr],
  \label{eq:pot}
\end{equation}
and the energy eigenvalue is $\Lambda$.

The wavenumber $\alpha$ is not a free parameter but is determined
by the magnetic geometry.
In Boozer coordinates $(\theta_B, \phi_B)$ the field-strength harmonic
$B_{m,n}$ varies as $\cos(m\theta_B - n\phi_B)$.
Along a field line, $\theta_B$ and $\phi_B$ are related by
$d\theta_B/d\phi_B = \iota$, the rotational transform, so
$B_{m,n}$ contributes a term with field-line wavenumber
\begin{equation}
  k_{m,n} = m - n/\iota.
  \label{eq:kmn}
\end{equation}
Here $n$ is the full toroidal mode number. When harmonics are tabulated
per field period, as is common for a device with $N_{\rm fp}$ periods, the
tabulated index $\bar n$ must be converted, $n = N_{\rm fp}\bar n$, before
Eq.~\eqref{eq:kmn} is applied.
The two dominant oscillatory harmonics of $B(\theta)$ along the
field line have wavenumbers $k_1 = m_1 - n_1/\iota$ and
$k_2 = m_2 - n_2/\iota$, so their ratio is
\begin{equation}
  \alpha = \frac{k_2}{k_1}
         = \frac{m_2 - n_2/\iota}{m_1 - n_1/\iota}.
  \label{eq:alpha}
\end{equation}
This is \emph{irrational} whenever $\iota$ is irrational and the two
modes are not in the same harmonic family ($m_1 n_2 \neq m_2 n_1$).
The Kolmogorov--Arnold--Moser (KAM) theorem guarantees that
these irrational tori persist under small perturbations of an
integrable magnetic field~\cite{ArnoldKAM}.

A remark on what is and is not aperiodic is in order. In a modular device the equilibrium is strictly
periodic in the Boozer angles, and the geometric coefficients of the ITG
equation are periodic functions of $(\theta_B,\phi_B)$. One may therefore
Fourier analyze in those angles and obtain a periodic problem with coupled
harmonics. The aperiodicity discussed here is a property of the
field-following coordinate, which is the natural coordinate for the
ballooning representation, and it is a property of the \emph{single}
harmonic content seen along one line. By Eq.~\eqref{eq:kmn} the ratio of
two field-line wavenumbers is irrational whenever $\iota$ is irrational,
and the $N_{\rm fp}$-fold module structure does not restore periodicity
along the line, because the line does not close. The two descriptions are
equivalent, with the localization discussed below manifest in the
field-following one. A corollary is that the two-harmonic potential of
Fig.~\ref{fig:bhatt_aah}(a) is the potential seen by a mode traveling along one field line, and
the full curvature spectrum contains many more harmonics than the two
retained here.

The introduction of the incommensurate term
$\lam\cos(\alpha\theta+\delta)$ is motivated at three distinct levels.
\emph{First}, it reflects physical reality: as shown in
Eqs.~\eqref{eq:kmn}--\eqref{eq:alpha}, the wavenumber ratio
$\alpha$ of the two dominant Boozer harmonics is determined by the
rotational transform $\iota$, and is irrational for almost all flux
surfaces. The amplitude $\lam$ is a measurable equilibrium property,
not a free parameter. Both $\alpha$ and $\lam$ can be extracted from an
equilibrium reconstruction independently of any transport
measurement.
\emph{Second}, it is what converts the Mathieu equation into the AAH
model: the second incommensurate harmonic (with $\alpha$ irrational)
destroys the band structure and replaces it with the sharp global
transition at $\lamAAH = 1$ --- it is precisely the irrationality of
$\alpha$ that makes the transition sharp, since a rational $\alpha$
would merely produce a longer-period Mathieu equation with bands.
\emph{Third}, near marginal stability the single-harmonic Mathieu
transition is a gradual crossover, not a sharp phase transition. The
incommensurate perturbation present naturally in the stellarator
equilibrium converts this crossover into the AAH transition.

\subsection{Domain of validity: magnetic shear and competing
localization mechanisms}
\label{ssec:shear}

The reduced equation was obtained in the long-wavelength limit
$k_\perp^2\rho_s^2 \to 0$. At finite global shear
\begin{equation}
  \hat s \;\equiv\; \frac{r}{q}\frac{dq}{dr}
        \;=\; -\,\frac{r}{\iota}\frac{d\iota}{dr},
  \label{eq:shat}
\end{equation}
the perpendicular wavenumber grows secularly along the field line,
$k_\perp^2 \simeq k_y^2[1 + (\hat s\theta)^2]$, and the resulting
finite-Larmor-radius damping confines the mode to
$|\theta_s| \sim 1/(\hat s\,k_y\rho_s)$ independently of any aperiodicity.
The Anderson mechanism described here confines it instead to
$2\pi\xiL$ in the same variable. The two are comparable when
\begin{equation}
  \hat s \;\sim\; \frac{1}{2\pi\,\xiL\,k_y\rho_s},
  \label{eq:shearcrossover}
\end{equation}
which for $\xiL \approx 3$ field-line periods and $k_y\rho_s \approx 0.5$
gives $\hat s \approx 0.1$. The AAH mechanism can therefore be expected to
dominate only at global shear well below this value. W7-X, whose global
shear is of the order of a few per cent, sits near rather than far inside
that bound, so the two effects are of comparable strength there rather
than cleanly separated.

In a low-shear device the local shear can spike to values far exceeding 
the global average, and such spikes are themselves an efficient mechanism for localizing ITG modes
along the field line~\cite{Plunk2014}. Estimating where the two compete
requires the full geometry --- the local shear enters the same coefficient
as the curvature and cannot be represented by the two-harmonic model used
here.

Two further mechanisms compete near the ends of the range in $\etai$.
Close to marginal stability, Landau damping and the kinetic drift
resonances omitted from the fluid closure supply their own damping along
the field line. Far from marginality, at large $\etai$, the single-well
fluid localization discussed in Section~\ref{sec:clamp} is strong and
is robustly observed in gyrokinetic computation. The window in which the
global aperiodic mechanism is the dominant one is therefore bounded from
both sides, and whether it is effective for a given device is a
quantitative question.

\subsection{The AAH exact result}
\label{ssec:aah}

The effective AAH coupling of the quasiperiodic component is
\begin{equation}
\lambda_{\rm AAH} = \frac{A\lambda F_\alpha}{2|t|}=
    \frac{\epsilon_n(1+\eta_i)\lam\mathcal{F}_\alpha}{2|t|},
\end{equation}
where $t$ and $\mathcal{F}_\alpha$ are the Wannier hopping and form
factors defined in Appendix~B.

Aubry and Andr\'{e}~\cite{Aubry1980} solved the discrete map to which
Eq.~\eqref{eq:bhatt} reduces in the tight-binding limit exactly. The result is:
\begin{itemize}
\item \(|\lambda_{\rm AAH}|<1\): extended states in the discrete AAH model.
\item \(|\lambda_{\rm AAH}|=1\): the critical point.
\item \(|\lambda_{\rm AAH}|>1\): exponentially localized states, with
\begin{equation}
    \xi_{\rm AAH}=\frac{1}{\ln|\lambda_{\rm AAH}|}.
      \label{eq:xi}
    \end{equation}
\end{itemize}
Here $\lam_{\rm AAH}$ is the dimensionless coupling of the
discrete tight-binding model (Appendix~B).

The result~\eqref{eq:xi} is exact for the \emph{discrete} AAH model,
requires no perturbation theory,
and holds for all eigenstates simultaneously and for all irrational
$\alpha$ (Fig.~\ref{fig:bhatt_aah}).
It is one of the rare exactly solvable problems in the theory of
disordered systems.

\begin{figure*}[!t]
\centering
\includegraphics[width=0.74\textwidth]{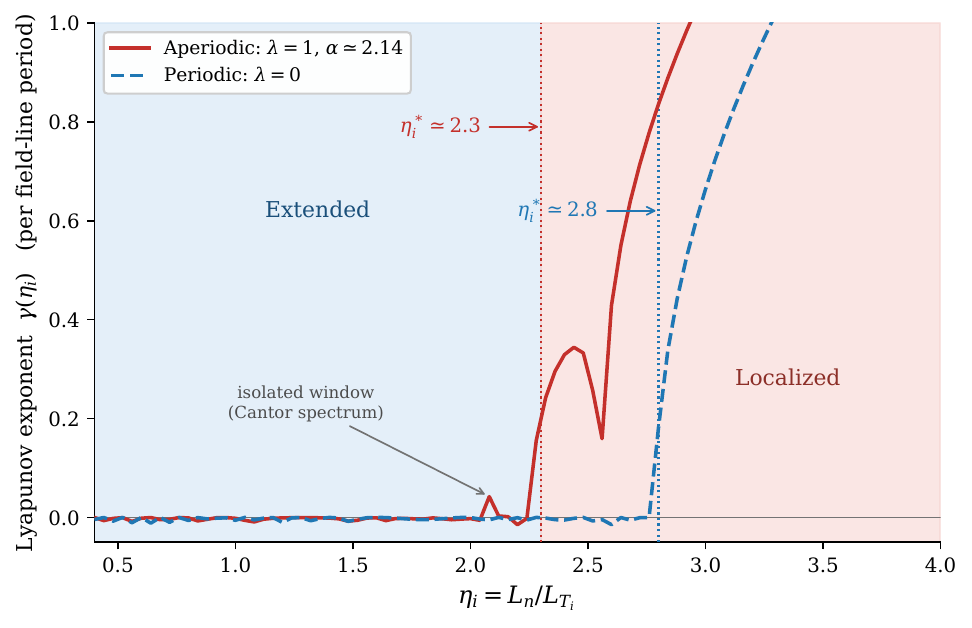}
\caption{\textbf{Localization threshold from the Lyapunov exponent.}
Lyapunov exponent \(\gamma(\eta_i)\) per
field-line period evaluated at \(\Lambda=0\) as a function of the ion
temperature gradient \(\eta_i\). The aperiodic model uses
\(\lambda=1\), \(\alpha=2+\sqrt{2}/10\simeq2.14\), and
\(\epsilon_n=0.12\) (the legend rounds this to \(2.1\)).
The periodic comparison has \(\lambda=0\) with the same
\(\epsilon_n\). The Lyapunov exponent becomes persistently positive at
\(\eta_i^*\simeq2.3\) for the aperiodic case and at
\(\eta_i^*\simeq2.8\) for the periodic comparison. For the periodic
case, \(\gamma>0\) indicates an ordinary Floquet gap rather than
Anderson localization. The exponent is converged: changing \(N\) from
\(200\) to \(3200\), or the integrator tolerance from \(10^{-6}\) to
\(10^{-9}\), leaves \(\gamma\) unchanged to better than \(10^{-3}\)
wherever \(\gamma>0.01\), and \(\gamma\) is independent of the phase
\(\delta\) to within \(0.5\%\). The residual structure near onset --- an
isolated positive window near \(\eta_i\approx2.08\), and a return to
\(\gamma\approx0\) at \(\eta_i\approx2.2\) --- is physical, reflecting the
Cantor structure of the spectrum, and is not numerical noise. The blue shaded region is the extended
mode-structure regime; the red shaded region is the localized
mode-structure regime, where the transport consequence is undetermined
in sign (Section~\ref{ssec:transport}).}
\label{fig:lyapunov}
\end{figure*}

\section{The localization threshold}
\label{sec:clamp}

\subsection{The Lyapunov exponent as the localization criterion}
\label{ssec:disc}

The AAH exact result (Section~\ref{ssec:aah}) establishes the
\emph{topology} of the localization transition: for irrational $\alpha$,
all eigenstates localize simultaneously at $\lam_{\rm AAH}=1$.
For the \emph{continuous} quasiperiodic Hill equation~\eqref{eq:bhatt},
the precise threshold must be computed from the equation itself, as done in ~\cite{Bhatt2026} for ballooning modes. To do so,
we write Eq.~\eqref{eq:bhatt} as a first-order system and integrate it
numerically along the field line. The fundamental quantity is the Lyapunov exponent.
Let \(T_N(\Lambda,\eta_i)\) be the transfer matrix that advances
\((\phi,\phi')\) from \(\theta=0\) to \(\theta=2\pi N\). The Lyapunov
exponent per field-line period is
\begin{equation}
    \gamma(\Lambda,\eta_i)
    =
    \lim_{N\to\infty}
    \frac{1}{N}\ln\|T_N(\Lambda,\eta_i)\|.
\end{equation}
In numerical work this limit is evaluated by accumulating the logarithms
of the norms after each period~\cite{Bennettin1980}. By the Oseledets theorem~\cite{Oseledets1968}, $\gamma$ exists and
is independent of the initial condition (for almost every initial
condition). A positive Lyapunov exponent indicates exponential separation of
solutions and the existence of exponentially decaying solutions when
the corresponding energy lies in the spectrum. We therefore use the
onset of \(\gamma>0\) at the drift-wave-resonant value \(\Lambda=0\)
as the localization diagnostic for the reduced model.
 A positive
exponent at a given $\Lambda$ means either that $\Lambda$ lies in the
spectrum and the corresponding state is exponentially localized, or that
$\Lambda$ lies in a spectral gap, where there is no state at all. In a
periodic device the second alternative is the only one available: the
positive exponent there is an ordinary Floquet band gap and is not
Anderson localization. For a quasiperiodic operator the gaps are dense and the
spectrum is a Cantor set~\cite{Avron1983}, so that as $A$ is increased at
fixed $\Lambda = 0$ the point $\Lambda = 0$ passes in and out of gaps.
The consequence, visible in the numerical results below, is that
$\gamma(\etai)$ is not monotonic near onset and $\etastar$ is properly
understood as the gradient above which a positive exponent becomes
\emph{generic}, not as a single sharp crossing. For the discrete AAH model, $\gamma = \ln\lam_{\rm AAH}$ for
$\lam_{\rm AAH} > 1$~\cite{Aubry1980}, confirming the sharp simultaneous
transition. The threshold $\etastar$ is the value of $\etai$ at which $\gamma$
becomes persistently positive. At $\Lambda = 0$, the potential amplitude $\epn(1+\etai)$ increases with $\etai$, driving the quasiperiodic
potential deeper until the Lyapunov exponent turns on, and modes are Anderson-localized.

\textbf{Numerical result.}
Figure~\ref{fig:lyapunov} shows $\gamma(\etai)$ computed with $N = 200$ field-line periods for the model
geometry ($\lam = 1$, $\alpha = 2+\sqrt{2}/10 \simeq 2.14$, $\epn = 0.12$) and
for the periodic device ($\lam = 0$, same $\epn$).
The Lyapunov exponent is noisy near the transition --- a signature
of the Cantor-set spectral structure of quasiperiodic
operators~\cite{Avron1983} --- but becomes clearly and persistently
positive above $\etastar \approx 2.3$ for the aperiodic case, and $\etastar \approx 2.8$ for the periodic case,
a factor of approximately $1.25\times$ lower threshold in the
aperiodic case.
The localization length $\xiL = 1/\gamma$ diverges as
$\etai\to\etastar{}$ from above and decreases for $\etai > \etastar$.

\textbf{Convergence and robustness.} Three checks are reported: \emph{(i)} Integration: the transfer matrix over each period
is obtained with an adaptive eighth-order Runge--Kutta scheme at relative
tolerance $10^{-9}$; reducing the tolerance to $10^{-6}$ changes $\gamma$
in the eighth significant figure. \emph{(ii)} Number of periods:
increasing $N$ from $200$ to $800$ and then to $3200$ changes $\gamma$ by
less than $10^{-3}$ at every $\etai$ for which $\gamma > 0.01$. The
residual scatter at the level of a few times $10^{-3}$ where $\gamma$
vanishes is the expected $O(N^{-1})$ sampling error and is not reduced by
tighter integration. \emph{(iii)} Phase: $\gamma$ is independent of the
relative phase $\delta$ of the incommensurate harmonic to within $0.5\%$
(for example $\gamma = 0.427$, $0.429$, $0.430$ and $0.427$ at
$\etai = 2.6$ for $\delta = 0$, $1$, $2$ and $3$), as required by the
Oseledets theorem.

The non-monotonicity anticipated above is visible in the data. Isolated
narrow windows of positive $\gamma$ occur below the persistent onset, for
instance a window of width $\Delta\etai \lesssim 0.04$ near
$\etai \approx 2.08$ with $\gamma \approx 0.04$, while $\gamma$ returns to
zero at $\etai = 2.16$--$2.24$ before rising steeply and remaining
positive from $\etai \approx 2.3$ onwards. The quoted
$\etastar \approx 2.3$ should accordingly be read as the edge of the
region in which localization is generic, with an uncertainty of order
$0.1$, and not as a sharp transition point.

\textbf{On the ordering of thresholds.} The ordering $\etailin < \etastar$ holds for the parameters
used here, but is not guaranteed. Both thresholds are set by the
same equilibrium quantities, and the localization threshold falls as the
incommensurate wavenumber $\alpha$ approaches a low integer, since the
quasi-periodic harmonic is then nearly resonant with a low Mathieu band. For
a curvature spectrum with small $|\alpha - N|$ at low $N$ one finds
$\etastar$ below the linear threshold, so that the mode
would localize before it became unstable.
This comparison carries a caveat of its own: $\etastar$ is computed here
from a reduced fluid mode-structure equation, whereas $\etailin$ comes from
a kinetic calculation~\cite{Zocco2018}, so the two are not outputs of a
single framework. Whether $\etailin < \etastar$
holds is therefore a property of a given equilibrium, and establishing it
securely would require both thresholds to be computed consistently.


It is important to distinguish the AAH localization from the
single-well localization that arises even in a periodic geometry.
In a single bad-curvature well, ITG eigenmodes acquire a Gaussian
envelope with localization parameter
$\lambda_{\rm loc} \sim \sqrt{\omega\bar\omega_d/\omega_t^2}$,
confining the mode within one connection length~\cite{HahmTang1988,RodriguezZocco2025}.
The AAH localization is qualitatively different: it arises from the
incommensurate aperiodicity across \emph{many} field-line periods,
with localization length $\xiL = 1/\gamma$.
Near marginal stability, single-well localization weakens
($\lambda_{\rm loc} \to 0$ as $\omega \to 0$~\cite{RodriguezZocco2025}),
and it is the global AAH mechanism that determines $\etastar$.

\subsection{Where the mode localizes}
\label{ssec:where}

It is natural to ask, and relevant experimentally, where along the field
line the localized mode sits. In the AAH model the center of a localized
state is determined by the realization of the quasiperiodic potential,
which here means by the phase $\delta$ of the incommensurate harmonic ---
that is, by the field-line label on the flux surface. There is accordingly
no preferred location on a given surface: different field lines localize
the mode at different positions, and the ensemble over field lines is
statistically homogeneous, with the toroidal symmetry broken only in the
sense that any individual realization selects a center. This is in
principle observable. At fixed radius, the fluctuation amplitude in the
localized regime should vary with field-line label in a way that it does
not in the extended regime, and the variation should be reproducible for a
fixed equilibrium, since $\delta$ is set by the geometry and not by the
turbulence.

\subsection{What localization implies for transport}
\label{ssec:transport}

The Anderson transition at $\etastar$ is a statement about mode structure.
Below $\etailin$ no modes are unstable; between $\etailin$ and $\etastar$
modes are linearly unstable and extended in ballooning space; above
$\etastar$ they remain linearly unstable but are exponentially confined
along the field line. What this implies for cross-field transport is a
separate question, and we do not think it is settled. Arguments run in both
directions and we set them out in turn.

The argument for suppression is as follows. A mode confined to a few field-line periods cannot sustain a
perturbation coherent across many connection lengths, and coherence over
many connection lengths is what an efficient transporting eddy requires.

The argument for the opposite conclusion is the standard ballooning
correspondence. A mode extended over $\Delta\theta$ in the ballooning
coordinate maps to a radial structure of width
$\Delta x \sim 1/(k_y \hat s\,\Delta\theta)$. Extended modes are
therefore radially \emph{narrow} and, by a naive mixing-length estimate, the less
effective transporters. On that reading, localization in ballooning space
would imply radially broader eddies and \emph{more} transport, not less.
There is a piece of experimental evidence supporting this argument: rational,
low-shear surfaces in W7-X are associated with transport
barriers~\cite{Andreeva2022,Chaudhary2024}, and if irrationality were
essential to localization then breaking it should, on the suppression
picture, have made confinement worse rather than better.

Neither argument is decisive here. Settling the question requires
nonlinear gyrokinetic simulation on either side of $\etastar$ in a
weak-shear configuration, measuring the heat flux rather than the mode
structure. We leave this to future work.

This unresolved sign is also what prevents us from drawing any conclusion
about ion-temperature clamping. If localization suppressed transport, $\etastar$ would bound the gradient
from \emph{below}: the profile could not relax past $\etastar$ without
re-entering the extended phase. A lower bound on the gradient leaves the
central temperature free to rise as heating power is added, which is the
opposite of what is observed. Clamping requires a bound from above --- a
mechanism by which transport rises sharply once some gradient is exceeded
--- and the suppression reading supplies the reverse. The opposite reading,
in which localized modes are the more effective transporters, would supply a
bound of the right sense, but we have no basis for preferring it. The sign
of the transport consequence is therefore the pivot on which any clamping
interpretation would turn, and since we cannot determine it, we make no
claim that the transition identified here accounts for the observed
clamping. We note in addition that the clamping is reported to be largely
independent of magnetic configuration and is seen in an axisymmetric device
as well~\cite{Beurskens2021,Beurskens2022}, which no mechanism resting on
3D geometry can by itself explain.

An analogy with the second stability regime of MHD ballooning theory
should be read in the same spirit. It is an analogy of structure --- a
second threshold above the first, beyond which the character of the
eigenmode changes --- and not a claim that the transport consequences are
analogous. In MHD the second boundary arises from genuine linear
restabilization by the Shafranov shift, and the modes above it are stable.
Here the modes above $\etastar$ remain unstable and only their spatial
structure has changed.

\subsection{Further predictions}
\label{ssec:Ti}

Three further qualitative predictions follow from the present model.
\emph{First}, $\etastar$ increases as $\lam$ decreases.
\emph{Second}, $\etastar$ increases as $\epn$ decreases: since the
potential depth is set by $A = \epn(1+\etai)$, a smaller $\epn$
(a more peaked density profile, i.e.\ smaller $\Ln$ at fixed $R$)
deepens the effective potential more slowly, requiring
higher $\etai$ to enter the localized phase.
\emph{Third}, deliberately changing $\lam$ --- for instance by
modifying the aperiodic harmonic content of the equilibrium field ---
should enable one to control $\etastar$ in a predictable, geometry-dependent way.
The quantitative implementation of such a perturbation in stellarator reactor design and experiments is left for
future work.

\section{Discussion}
\label{sec:disc}

The argument presented here is deliberately minimal.
The AAH mapping identifies a structural property of the
ITG eigenvalue problem in aperiodic geometry --- the existence of a
sharp localization transition --- and that property, rather than any transport or
profile consequence, is the result of this paper.

Several limitations should be stated clearly.
\emph{First}, the connection between AAH localization in ballooning
space and cross-field transport is not established, and not even
established in sign, as discussed in
Section~\ref{ssec:transport}. A quantitative derivation, or failing that a
nonlinear gyrokinetic measurement of the heat flux on either side of
$\etastar$ in a weak-shear configuration, is the natural next step.
\emph{Second}, the present model is applicable in a fluid, long-wavelength
regime. Electromagnetic effects, trapped electrons, and finite Larmor radius
corrections will modify the coefficients in the fluid equations. 
The kinetic framework (for instance,~\cite{RodriguezZocco2025})
provides the appropriate description of single-well localization
with these effects retained; incorporating their kinetic dispersion
function into the global aperiodic eigenvalue problem
is the natural next step.
\emph{Third}, the AAH exact results ($\gamma = \ln\lam_{\rm AAH}$,
transition at $\lam_{\rm AAH}=1$) apply to the \emph{discrete}
tight-binding model with effective coupling $\lam_{\rm AAH}$
(Appendix~B), not to the geometric ratio
$\lam$ directly. The present equation is a \emph{continuous} quasiperiodic Hill
equation, for which the localization transition is generic
but the precise threshold is sensitive to the precise magnetic geometry.
\emph{Fourth}, the analysis is carried out on the drift-wave-resonant
branch and not on the ITG branch proper (Section~\ref{ssec:bhatt}), and it
neglects magnetic shear, which restricts it to weak-shear devices and
leaves its competition with local-shear localization unquantified
(Section~\ref{ssec:shear}).

We remark on two further issues.
The present model is linear, and says nothing about zonal flows. How a
population of modes localized along the field line couples to zonal
structures, and whether localization alters the zonal-flow damping that
regulates ITG saturation, cannot be addressed within it. Second, a
Ginzburg--Landau formulation for the aperiodic operator, in the spirit of
the treatment given for the ballooning network in
Ref.~\cite{Bhatt2026}, would be a natural route to the nonlinear question
and we regard it as the most promising analytic continuation of this work.
Bhattacharjee~\cite{Bhatt2026} shows that geometric aperiodicity
localizes MHD pressure-driven eigenmodes and formulates the resulting
nonlinear stability as a connectivity phase transition on a
Ginzburg--Landau network. The present paper identifies the same aperiodicity-driven localization
for ITG eigenmodes. The structural parallel with the MHD
second stability boundary is close, but the transport parallel needs to be considered carefully. The two mechanisms are independent --- MHD ballooning stability and
ITG transport operate on different timescales and involve different
mode types --- but both arise from the same underlying structure:
aperiodic coefficients in the ballooning equation producing
Anderson-localized eigenmodes.

\section{Conclusion}
\label{sec:conc}

We have established the following:
\begin{enumerate}\setlength\itemsep{2pt}
  \item The ITG eigenvalue equation in aperiodic stellarator geometry
    is a quasiperiodic Hill equation whose structure maps onto the
    Aubry--Andr\'{e}--Harper model in the limit of a single dominant
    incommensurate curvature harmonic and long perpendicular wavelength.
    The reduction to a quasiperiodic Hill equation is rigorous within the
    fluid closure. The further reduction to an AAH difference equation is
    a controlled tight-binding approximation.
  \item For the reduced continuous Hill equation, the relevant localization
  diagnostic is the Lyapunov exponent. In the illustrative two-harmonic
  model, the Lyapunov exponent at the drift-wave-resonant value
  \(\Lambda=0\) becomes positive above a geometry-dependent threshold
  \(\eta_i^*\). This is the continuous-equation analogue of the sharp
  AAH transition in the tight-binding limit. A periodic device also
  develops a positive exponent, but there it is an ordinary Floquet band
  gap and not Anderson localization.
  \item For the model two-harmonic curvature profile
    $K(\theta) = \cos\theta + \lam\cos(\alpha\theta)$ with
    $\lam = 1$, $\alpha \simeq 2.14$, $\epn = 0.12$,
    the Lyapunov exponent gives the gradient threshold $\etastar \approx 2.3$,
    against $\approx 2.8$ for the periodic comparison. Both values carry an
    uncertainty of order $0.1$ arising from the Cantor structure of the
    spectrum rather than from numerical convergence, which was verified
    against integrator tolerance, the number of field-line periods, and the
    phase of the incommensurate harmonic.
    The accurate threshold for any real device requires computing
    the Lyapunov exponent from the full Boozer harmonic spectrum of
    the equilibrium state for the device, which is sensitive to geometry.
  \item Above \(\eta_i^*\), the reduced eigenmode problem enters a regime with positive Lyapunov exponent and exponentially localized
   field-line structure. What this implies for cross-field
   heat transport is not determined by the present calculation: confinement of the mode along the field line
   argues for suppression, while the ballooning correspondence between
   parallel extent and radial width argues for enhancement, and that
   correspondence itself weakens in the weak-shear limit in which the
   mechanism operates. Nonlinear gyrokinetic simulation on either side of
   \(\eta_i^*\) in a weak-shear configuration is required to settle it.
\end{enumerate}

\section*{Appendix A: From the fluid ITG ballooning equation to the quasiperiodic Hill model}

We derive the reduced drift-wave-resonant ITG equation used in the
main text. The purpose of the derivation is not to compute the full
kinetic ITG growth rate, but to obtain the leading-order mode-structure
equation along a field line. The calculation follows the spirit of the
straight-stellarator fluid model of Bhattacharjee et al., but we write
the magnetic-drift drive in a form suitable for general stellarator
geometry.

Let \(s\) denote distance along a field line, \(x\) the radial coordinate,
and \(y\) the binormal coordinate. We consider perturbations of the form
\[
\delta f = \tilde f(s)\exp(i k_y y-i\omega t).
\]
The equilibrium density and ion temperature satisfy
\[
\frac{d\ln n_0}{dx}=-\frac{1}{L_n},\qquad
\frac{d\ln T_{i0}}{dx}=-\frac{1}{L_{T_i}},
\qquad
\eta_i\equiv \frac{L_n}{L_{T_i}} .
\]
We define
\[
\varphi \equiv \frac{e\delta\Phi}{T_e},\qquad
N\equiv \frac{\delta n_i}{n_0},\qquad
\Theta\equiv \frac{\delta T_i}{T_i},
\]
and
\[
\omega_*\equiv \omega_{*e}
=
\frac{k_y c T_e}{eB L_n}.
\]
The field-line-dependent magnetic-drift frequency is denoted
\(\omega_d(s)\). We define the dimensionless drift function
\[
\epsilon_d(s)\equiv \frac{\omega_d(s)}{\omega_*}.
\]
In a simple large-aspect-ratio circular tokamak model,
\(\epsilon_d\simeq 2(L_n/R)\cos\theta\), up to sign convention; the
factor of two is absorbed into \(K(\theta)\) by the normalization
convention of Eq.~\eqref{eq:omegad}, which fixes \(|K|\sim 1\).
In a stellarator, however, \(\omega_d(s)\) should be computed from the
actual curvature and grad-\(B\) drift projected onto the binormal
wavevector.

For adiabatic electrons,
\begin{equation}
\frac{\delta n_e}{n_0}=\frac{e\delta\Phi}{T_e}=\varphi .
\end{equation}
In the long-wavelength limit \(k_\perp^2\rho_s^2\ll 1\), quasineutrality
then gives
\begin{equation}
N=\varphi .
\end{equation}

The ion temperature perturbation is produced by \(E\times B\) advection
of the equilibrium ion-temperature gradient:
\begin{equation}
\frac{\partial \Theta}{\partial t}
+
\mathbf v_E\cdot \nabla\ln T_i=0 .
\end{equation}
Since
\[
v_{E x}=-\frac{c}{B}\frac{\partial \delta\Phi}{\partial y}
=
-i k_y\frac{c}{B}\delta\Phi ,
\]
we obtain
\[
-i\omega \Theta+i\eta_i\omega_*\varphi=0,
\]
or
\[
\Theta=\frac{\eta_i\omega_*}{\omega}\varphi .
\]

The linearized ion continuity equation is
\begin{equation}
-i\omega N+i\omega_*\varphi
+
\frac{\partial u_\parallel}{\partial s}
-
i\omega_d(s)(N+\Theta)=0 .
\end{equation}
The sign convention has been chosen such that positive \(\omega_d\)
corresponds to bad curvature and gives the destabilizing sign in the
final Hill equation. Using \(N=\varphi\), this becomes
\[
-i(\omega-\omega_*)\varphi
+
\frac{\partial u_\parallel}{\partial s}
-
i\omega_d(s)
\left(
1+\frac{\eta_i\omega_*}{\omega}
\right)\varphi
=0 .
\]

The minimal parallel ion momentum equation is
\[
-i\omega u_\parallel
=
-c_s^2\frac{\partial\varphi}{\partial s},
\qquad
c_s^2=\frac{T_e}{m_i}.
\]
Thus
\[
\frac{\partial u_\parallel}{\partial s}
=
-i\frac{c_s^2}{\omega}
\frac{\partial^2\varphi}{\partial s^2}.
\]
Substitution gives
\[
-i(\omega-\omega_*)\varphi
-
i\frac{c_s^2}{\omega}
\frac{\partial^2\varphi}{\partial s^2}
-
i\omega_d(s)
\left(
1+\frac{\eta_i\omega_*}{\omega}
\right)\varphi
=0 .
\]
Multiplying by \(i\omega\), we obtain the fluid equation
\begin{equation}
c_s^2\frac{\partial^2\varphi}{\partial s^2}
+
\left[
\omega(\omega-\omega_*)
+
\omega_d(s)(\omega+\eta_i\omega_*)
\right]\varphi
=0 .
\end{equation}

Now introduce
\[
\theta=\frac{\omega_*}{c_s}s,\qquad
\Omega=\frac{\omega}{\omega_*},\qquad
\epsilon_d(\theta)=\frac{\omega_d(\theta)}{\omega_*}.
\]
The dimensionless equation is
\[
\frac{d^2\varphi}{d\theta^2}
+
\left[
\Omega(\Omega-1)
+
\epsilon_d(\theta)(\Omega+\eta_i)
\right]\varphi
=0 .
\]
If we write \(\epsilon_d(\theta)=\epsilon_n K(\theta)\), this becomes
\[
\frac{d^2\varphi}{d\theta^2}
+
\left[
\Omega(\Omega-1)
+
\epsilon_n K(\theta)(\Omega+\eta_i)
\right]\varphi
=0 .
\]
This is equation (1), our point of departure in the main text.

\section*{Appendix B: From the continuous Hill equation to the tight-binding AAH model}
\label{app:wannier}

We derive the explicit form of the hopping amplitude $t$ and
on-site modulation $V$ that connect the continuous quasiperiodic
Hill equation~\eqref{eq:bhatt} to the discrete
Aubry--Andr\'{e}--Harper tight-binding model, via Wannier function
projection onto the lowest Mathieu band.

We set $\lam = 0$ and write Eq.~\eqref{eq:bhatt} in Schr\"{o}dinger
form:
\begin{equation}
    \hat H_0\psi
    \equiv
    \left(
        -\frac{d^2}{d\theta^2}
        -
        A\cos\theta
    \right)\psi
    =
    E\psi,
    \qquad
    A=\epsilon_n(1+\eta_i).
  \label{eq:schrodinger_periodic}
\end{equation}
with $A = \epn(1+\etai)$.
This is the Mathieu equation, whose solutions are Bloch functions
$\psi_{k}(\theta) = e^{ik\theta}\,u_k(\theta)$ with $u_k$ periodic
and band energies $E_0(k)$ for the lowest band.

The Wannier function for the lowest band centerd on site $m$
(field-line period $2\pi m$) is
\begin{equation}
  w_m(\theta) = w_0(\theta - 2\pi m)
  = \frac{1}{2\pi}\int_0^1 e^{-2\pi ikm}\,\psi_k(\theta)\,dk,
  \label{eq:wannier_def}
\end{equation}
where the integral runs over the first Brillouin zone $k\in[0,1)$.
These are exponentially localized around each curvature well
and satisfy $\langle w_m | w_{m'}\rangle = \delta_{mm'}$.

Projecting $\hat{H}_0$ onto the Wannier basis gives the tight-binding
matrix elements.
The nearest-neighbor hopping amplitude is
\begin{equation}
  t \;\equiv\; -\langle w_0 \,|\, \hat{H}_0 \,|\, w_1 \rangle
  \;=\; -\!\int_{-\infty}^{\infty}\!
        w_0(\theta)\,\hat{H}_0\, w_0(\theta - 2\pi)\,d\theta.
  \label{eq:t_def}
\end{equation}
Since $\hat{H}_0$ is diagonal in the Bloch basis with eigenvalue $E_0(k)$,
the Wannier matrix element~\eqref{eq:t_def} is the first Fourier
coefficient of $E_0(k)$:
\begin{equation}
  t = -\int_0^1 E_0(k)\,e^{2\pi ik}\,dk
  \;\approx\; \frac{E_0(k=0) - E_0(k=\tfrac{1}{2})}{4},
  \label{eq:t_bandstructure}
\end{equation}
where the approximation retains only the fundamental harmonic of
$E_0(k)$ (justified when higher harmonics $t_2, t_3, \ldots$ are
exponentially small in $A$).
Equation~\eqref{eq:t_bandstructure} identifies $t$ as one
quarter of the lowest Mathieu band's full width,
$t = \tfrac{1}{4}\,W(A)$, with $W = E_0(0) - E_0(\tfrac{1}{2})$.

The quasiperiodic part $A\lam\cos(\alpha\theta)$ gives
diagonal matrix elements
\begin{equation}
  V_m \;\equiv\; A\lam\,\langle w_m\,|\,\cos(\alpha\theta)\,|\,w_m\rangle
  \;=\; A\lam\,\mathcal{F}_\alpha\,\cos(2\pi\alpha m + \delta),
  \label{eq:Vm}
\end{equation}
where the form factor
\begin{equation}
  \mathcal{F}_\alpha
  \;\equiv\; \int_{-\infty}^{\infty}
             |w_0(\theta)|^2\,\cos(\alpha\theta)\,d\theta
  \label{eq:Falpha}
\end{equation}
is the cosine transform of the Wannier probability density at
the quasiperiodic wavenumber $\alpha$.
For small $A$, $w_0(\theta)$ is approximately a Gaussian of width
$\sigma \sim (A/2)^{-1/4}$, giving
$\mathcal{F}_\alpha \approx \exp(-\alpha^2\sigma^2/2)$;
for larger $A$ the exact value must be computed numerically from
the Mathieu eigenfunctions.

Off-diagonal matrix elements $\langle w_m|\cos(\alpha\theta)|w_{m'}\rangle$
with $m \neq m'$ are suppressed by the exponential localization of
the Wannier functions and are neglected in the single-band approximation.

Expanding $\phi(\theta) = \sum_m c_m\,w_0(\theta - 2\pi m)$ and
retaining only nearest-neighbor hopping, the eigenvalue problem
reduces to
\begin{equation}
  -t\,(c_{m+1} + c_{m-1})
  + A\lam\,\mathcal{F}_\alpha\,\cos(2\pi\alpha m + \delta)\,c_m
  = (E - E_0)\,c_m,
  \label{eq:tb_aah}
\end{equation}
which is the standard Aubry--Andr\'{e}--Harper equation~\cite{Aubry1980}
with coupling constant

\begin{equation}
    \lambda_{\rm AAH} = \frac{A\lambda F_\alpha}{2|t|}
    =
    \frac{\epsilon_n(1+\eta_i)\lambda F_\alpha}{2|t|}.
\end{equation}
The Aubry--Andr\'{e} transition occurs at
\begin{equation}
    |\lambda_{\rm AAH}|=1,
    \qquad
    \text{or equivalently}
    \qquad
    A\lambda F_\alpha=2|t|.
\end{equation}

The Aubry--Andr\'{e} exact result gives localization for
$\lambda_{\rm AAH} > 1$,
which is a transcendental condition on $A = \epn(1+\etai)$
that determines the single-band estimate of the threshold $\etastar$.
Here $W(A) = E_0(k=0) - E_0(k=\tfrac{1}{2})$.
The exact threshold for the \emph{continuous} equation --- used
throughout the main text --- is given instead by the Mathieu
Lyapunov condition $\gamma(\etastar) = 0$ (Section~\ref{ssec:disc}),
which automatically includes inter-band coupling and higher-order
hoppings.
A quantitative comparison of the discrete ($\lam_{\rm AAH} = 1$) and
continuous ($\gamma = 0$) thresholds requires computing $W(A)$ and
$\mathcal{F}_\alpha(A)$ from the actual Mathieu eigenfunctions, which
we leave to future work.

\section*{Acknowledgements}
We would like to thank Per Helander and Hongxuan Zhu for enlightening discussions. We especially thank Per Helander and the astute reviewers of an earlier version of this paper for trenchant criticism that greatly improved the paper. We acknowledge the use of ChatGPT and Claude (Anthropic) for writing assistance, crafting illustrations, and literature search. This research is supported by the Department of Energy Award No. DE-SC0024548. 

\section*{Competing Interests}
We declare no competing interests.

\section*{Data Availability}
No new experimental data were generated.
The scripts used to compute the transfer matrices, the Lyapunov
exponents of Fig.~\ref{fig:lyapunov}, and the convergence and
phase-independence checks reported in Section~\ref{ssec:disc} are
available upon request.


\end{document}